\documentstyle[aps,preprint]{revtex}
\begin{document}
\draft
\title{Hartree-Fock treatment of the two-component Bose-Einstein condensate}
\author{Patrik \"Ohberg$^{1}$ and Stig Stenholm$^2$}
\address{$^1$Helsinki Institute of Physics}
\address{P. O. Box 9, FIN-00014 University of Helsinki}
\address{FINLAND}
\address{$^2$Department of Physics, Royal Institute of Technology}
\address{Lindstedtv\"agen 24, S-10044 Stockholm} 
\address{SWEDEN}

\date{\today}
\maketitle

\begin{abstract}
We present a numerical study of a trapped binary Bose-condensed gas by
solving the corresponding Hartree-Fock equations. The density profile
of the binary Bose gas is solved with a harmonic trapping potential as
a function of temperature in two and three dimensions. We find a symmetry 
breaking in the two dimensional case where the two 
condensates separate.
We also present a phase diagram in the three dimensional case of the 
different regions where the binary 
condensate becomes a single condensate and eventually an ordinary 
gas as function of temperature and the interaction strength 
between the atoms.
\end{abstract}


\section{Introduction}

The recent realization of Bose-Einstein condensation in dilute gases of 
trapped alkali atoms has provided an opportunity to investigate macroscopic 
quantum effects in novel systems \cite{Cor95,Hul95,Ket95}. 
The second generation of experiments on 
condensates in Rb 
and Na vapors \cite{jila,Mew96,And96,mit,Myat} give detailed information 
about the condensate long range order
 which 
distinguishes them from an ordinary gas. These atoms are known to have a 
positive scattering length , i.e the atoms experience a 
repulsive force between them.
 Recent experiments with Li \cite{lit} show that, 
for a limited range of particle densities 
even particles with negative scattering length can form a condensate. In a 
recent experiment one has been able to trap atoms of two different spin states
 and cool them below the condensate transition point \cite{Myat}. 
These experiments 
give us knowledge about the condensation in a whole knew dynamical regime. 

Much work on the density distribution of the condensate has been carried out 
for a single condensate and lately also for the two-component condensate 
\cite{KBG2,Griff1,Esry}. 
These 
calculations have mainly been valid at the temperatures $T=0$. In our earlier 
paper we have solved the Hartree-Fock equations 
for a single condensate \cite{us}. 
In this paper we derive and 
solve numerically the Hartree-Fock equations for a two-component gas 
trapped in 
harmonic external potentials at a finite temperature. We solve for the 
density in two and three 
dimensions for the case with simple harmonic potentials.
The Hartree-Fock equations give a phase transition at the onset of 
condensation but they do not take into account any critical fluctuations. 
It is however possible to obtain a qualitative picture of the features 
of the Bose-Einstein phase transition using the HF approach.

In a real experiment, the different potentials experienced by the two 
components make them respond differently to gravity. This separates the 
condensed clouds in space, but we assume this effect to be compensated by 
technical means. In this way we can investigate the intrinsic effects 
determined by the physically more essential interactions. 
The two components of the condensate repell each other, and if they form 
exactly on top of each other, the more weakly interacting one is 
pushed to expand 
away from the center. This is found to lead to a condensate component that 
has a minimum at its center. This situation is still not optimal, because 
energy can be lowered by separating the condensates. In a symmetric trap 
this implies a breaking of the symmetry; the two peaks are shifted in a 
direction not fixed by the external potentials. 
We investigate this symmetry breaking in a
 two-dimensional model trap, because solving for the asymmetric condensate 
proves to be 
numerically demanding. The calculations are found to verify the intuitive 
picture described.

If the two components did not interact, the condensates would form and 
disappear independently, thus showing two different transition temperatures. 
With condensate interactions, the presence of one condensate will affect 
the formation 
of the other one. This will shift the transition temperatures depending on the 
strength of the interaction between the two components. In particular, we 
expect a repulsive interaction to work against the simultaneous formation of 
both 
condensates. Thus we expect to see a modification of the phase diagram in the 
plane of temperature and interaction between the components. In order to keep 
the numerical effort manageable, we investigate this effect in a spherically 
symmetric three-dimensional trap. We find that 
the interaction shifts the boundary 
between the two- and one-condensate regions. We even find a case where there 
is no formation of a single-condensate region.

The organization of the paper is as follows: Section II reviews the HF 
equations that we solve numerically. In section III we briefly go through the 
numerical methods that have been used. The results of the calculations are 
presented in Sec. IV where various situations are compared. Finally Section V 
comments on the calculations and their results. 

\section{The Hartree-Fock equations for the two-component condensate}

The Hamiltonian for an interacting Bose gas of two different kinds of atoms 
can be written in the form
\begin{equation} \label{ham1}
\hat H =\hat H_{1}+\hat H_{2}+\hat H_{3},
\end{equation}
where
\begin{eqnarray}
\hat H_{1}&=&\int d{\bf r}\psi_{1}^{\dagger}({\bf r})[-{\hbar^2 \over {2m_{1}}}
\nabla^2+U_{1}({\bf r})] \psi_{1}({\bf r})
+{1\over 2} \int d{\bf r} \int d{\bf r'}
\psi_{1}^{\dagger}({\bf r'})\psi_{1}^{\dagger}({\bf r})V_{1}({\bf r}-{\bf r'})
\psi_{1}({\bf r})\psi_{1}({\bf r'}) \nonumber \\
\hat H_{2}&=& \int d{\bf r}\psi_{2}^{\dagger}({\bf r})[-{\hbar^2 \over 
{2m_{2}}}\nabla^2+U_{2}({\bf r})] \psi_{2}({\bf r})
+{1\over 2} \int d{\bf r} \int d{\bf r'}
\psi_{2}^{\dagger}({\bf r'})\psi_{2}^{\dagger}({\bf r})V_{2}({\bf r}-{\bf r'})
\psi_{2}({\bf r})\psi_{2}({\bf r'})  \\
\hat H_{3} &=& {1\over 2} \int d{\bf r} \int d{\bf r'}
\psi_{1}^{\dagger}({\bf r'})\psi_{2}^{\dagger}({\bf r})V_{int}
({\bf r}-{\bf r'})\psi_{1}({\bf r})\psi_{2}({\bf r'}) \nonumber
\end{eqnarray}
and $\psi_{1},\psi_{1}^{\dagger},\psi_{2}$ and $\psi_{2}^{\dagger}$ are the 
Boson field operators which obey the commutation rule
\begin{equation}
[\psi_{i}({\bf r}),\psi_{j}^{\dagger}({\bf r'})]=\delta_{ij}\delta 
({\bf r}-{\bf r'}).
\end{equation}
$U_{1}({\bf r})$ and $U_{1}({\bf r})$ are the two different external traps 
that confine the atoms. We are here going to use the short range approximation 
both for the interaction between the particles of the same kind and between 
the two different kinds of particles. This gives us the interaction potentials
\begin{eqnarray}
V_{1}({\bf r}-{\bf r'})=v_{1} \delta({\bf r}-{\bf r'}) \\
V_{2}({\bf r}-{\bf r'})=v_{2} \delta({\bf r}-{\bf r'}) \\
V_{int}({\bf r}-{\bf r'})=w \delta({\bf r}-{\bf r'}),
\end{eqnarray}
where 
\begin{equation}
v_{1}={{4 \pi \hbar^2 a_{1}}\over m_{1}} \qquad v_{2}={{4 \pi \hbar^2 a_{2}}
\over m_{2}} \qquad w={{4 \pi \hbar^2 a_{12}}\over \sqrt{m_{1} m_{2}}}
\end{equation}
with the s-wave scattering lengths $a_{1},a_{2}$ and $a_{12}$. Our goal is to 
calculate the temperature dependence of the two gases. We therefore introduce 
the thermodynamic free energy $\Omega (T,\mu)$ which is defined as
\begin{equation}
e^{-\beta \Omega}=Tr [e^{-\beta (\hat H - \mu_{i} \hat N^{(i)})}],
\end{equation}
where $\{ \mu_{1},\mu_{2}\}$ are the chemical potentials and 
$\mu_{i} \hat N^{(i)}$ is a sum over the two different gases. 
We then use a thermodynamic variational principle
\begin{equation}
\Omega ({\hat H})\le\Omega ({\hat H^{t}})+\langle ({\hat H}-
{\hat H^{t}})\rangle_{t},
\end{equation}
where
\begin{equation}
\langle {\hat A} \rangle_{t} \equiv {{{Tr} [{e}^{\beta ({\hat H^{t}}-
\mu_{i} \hat{N^{(i)}})} {\hat A}]}\over {{Tr} [{e}^{\beta ({\hat H^{t}}-
\mu_{i} \hat{N^{(i)}})}]}}
\end{equation}
and $\hat H^{t}$ is a single particle trial Hamiltonian 
\begin{eqnarray}
\hat H^{t}&=&\hat H_{1}^{t}+\hat H_{2}^{t}=\sum_{\alpha} \{E_{\alpha}^{(1)}
a_{\alpha}^{\dagger}a_{\alpha}+E_{\alpha}^{(2)}b_{\alpha}^
{\dagger}b_{\alpha} \}  \nonumber \\ &=& 
\sum_{\alpha}\int d{\bf r} \{E_{\alpha}^{(1)} |\varphi_{\alpha}({\bf r})|^{2}
a_{\alpha}^{\dagger}a_{\alpha}+E_{\alpha}^{(2)}
|\phi_{\alpha}({\bf r})|^{2}b_{\alpha}^{\dagger}b_{\alpha} \}.
\end{eqnarray}
This gives for the thermodynamic potential
\begin{equation}
\Omega ({\hat H})\le \Omega ({\hat H_{1}^{t}})+\Omega ({\hat H_{2}^{t}})+
\langle {\hat H_{1}}-{\hat H_{1}^{t}}\rangle_{t}+
\langle {\hat H_{2}}-{\hat H_{2}^{t}}\rangle_{t}+
\langle {\hat H_{3}}\rangle_{t}.
\end{equation}
The single particle states $\varphi_{\alpha}$ and $\phi_{\alpha}$ are to be 
determined such that they minimize the thermodynamic free energy $\Omega$. 
We now expand our field operators
\begin{eqnarray}
\psi_{1}({\bf r})&=&\sum_{\alpha}\varphi_{\alpha}({\bf r}) a_{\alpha} \\
\psi_{1}^{\dagger}({\bf r})&=&\sum_{\alpha}\varphi_{\alpha}^{*}
({\bf r}) a_{\alpha}^{\dagger} \\
\psi_{2}({\bf r})&=&\sum_{\alpha}\phi_{\alpha}({\bf r}) b_{\alpha} \\
\psi_{2}^{\dagger}({\bf r})&=&\sum_{\alpha}\phi_{\alpha}^{*}
({\bf r}) b_{\alpha}^{\dagger}.
\end{eqnarray}
Inserting this expansion into Eq. \ref{ham1} gives the Hamiltonians
\begin{eqnarray}
\hat H_{1} &=& \int d{\bf r} \sum_{\alpha}\varphi_{\alpha}^{*}({\bf r})[
-{\hbar^2 \over {2m_{1}}}\nabla^2+U_{1}({\bf r})] \varphi_{\alpha}({\bf r})
a_{\alpha}^{\dagger}a_{\alpha} \nonumber \\ && \\
&& +{1\over 2} v_{1} \int d{\bf r} \sum_{\alpha\beta\gamma\delta}
\varphi_{\alpha}^{*}({\bf r})\varphi_{\beta}^{*}({\bf r})\varphi_{\gamma}
({\bf r})\varphi_{\delta}({\bf r})a_{\alpha}^{\dagger}a_{\beta}^{\dagger}
a_{\gamma}a_{\delta} \nonumber \\
&& \nonumber \\
\hat H_{2} &=& \int d{\bf r} \sum_{\alpha}\phi_{\alpha}^{*}({\bf r})[
-{\hbar^2 \over {2m_{2}}}\nabla^2+U_{2}({\bf r})] \phi_{\alpha}({\bf r})
b_{\alpha}^{\dagger}b_{\alpha} \nonumber \\ && \\
&& +{1\over 2} v_{1} \int d{\bf r} \sum_{\alpha\beta\gamma\delta}
\phi_{\alpha}^{*}({\bf r})\phi_{\beta}^{*}({\bf r})\phi_{\gamma}
({\bf r})\phi_{\delta}({\bf r})b_{\alpha}^{\dagger}b_{\beta}^{\dagger}
b_{\gamma}b_{\delta} \nonumber \\
&& \nonumber \\
\hat H_{3} &=& w \int d{\bf r} \sum_{\alpha\beta\gamma\delta}
\varphi_{\alpha}^{*}({\bf r})\phi_{\beta}^{*}({\bf r})
\phi_{\gamma}({\bf r})\varphi_{\delta}({\bf r})a_{\alpha}^{\dagger}
b_{\beta}^{\dagger}b_{\gamma}a_{\delta}.
\end{eqnarray}
We can now use the single particle Hamiltonians and rewrite the thermodynamic 
free energy as
\begin{equation}
\Omega=\Omega_{1}^{t}+\Omega_{2}^{t}+\int d{\bf r} \tilde \Omega({\bf r})
\end{equation}
where
\begin{eqnarray}
\Omega_{1}^{t}&=&-{1\over \beta}\sum_{\alpha} ln[1-e^{-\beta (E_{\alpha}^{(1)}-
\mu_{1})}] \\
\Omega_{2}^{t}&=&-{1\over \beta}\sum_{\alpha} ln[1-e^{-\beta (E_{\alpha}^{(2)}-
\mu_{2})}] \\
\end{eqnarray}
and $\tilde \Omega =\tilde \Omega_{1}+\tilde \Omega_{2}+
\tilde \Omega_{3}$ with 
\begin{eqnarray}
\tilde \Omega_{1} ({\bf r})&=& \sum_{\alpha}\varphi_{\alpha}^{*}({\bf r})[
-{\hbar^2 \over {2m_{1}}}\nabla^2+U_{1}({\bf r})-E_{\alpha}^{(2)}] 
\varphi_{\alpha}({\bf r}) \langle N_{\alpha}^{(1)}\rangle_{t} \nonumber \\
&& \\
&& +{v_{1}\over 2}\sum_{\alpha} |\varphi_{\alpha}
({\bf r})|^{4}\langle N_{\alpha}^{(1)}(N_{\alpha}^{(1)}-1) \rangle_{t}+
 v_{1} \sum_{\alpha \ne \beta}|\varphi_{\alpha}({\bf r})|^{2}|\varphi_{\beta}
({\bf r})|^{2}\langle N_{\alpha}^{(1)}N_{\beta}^{(1)}\rangle_{t} \nonumber \\
&& \nonumber \\
\tilde \Omega_{2} ({\bf r})&=& \sum_{\alpha}\phi_{\alpha}^{*}({\bf r})[
-{\hbar^2 \over {2m_{2}}}\nabla^2+U_{2}({\bf r})-E_{\alpha}^{(2)}] 
\phi_{\alpha}({\bf r}) \langle N_{\alpha}^{(2)}\rangle_{t} \nonumber \\
&& \\
&& +{v_{2}\over 2}\sum_{\alpha} |\phi_{\alpha}
({\bf r})|^{4}\langle N_{\alpha}^{(2)}(N_{\alpha}^{(2)}-1) \rangle_{t}+
 v_{2} \sum_{\alpha \ne \beta}|\phi_{\alpha}({\bf r})|^{2}|\phi_{\beta}
({\bf r})|^{2}\langle N_{\alpha}^{(2)}N_{\beta}^{(2)}\rangle_{t} \nonumber \\
&& \nonumber \\
\tilde \Omega_{3}({\bf r}) &=& w \sum_{\alpha\beta} |\varphi_{\alpha}({\bf r})
|^{2}|\phi_{\beta}({\bf r})|^{2}\langle N_{\alpha}^{(1)}N_{\beta}^{(2)}
\rangle_{t}.
\end{eqnarray}
Here we have used the independent particle properties of $\hat H_{1}^{t}$ and 
$\hat H_{2}^{t}$ with
\begin{eqnarray} 
N_{\alpha}^{(1)}&=& a_{\alpha}^{\dagger}a_{\alpha} \\
N_{\alpha}^{(2)}&=& b_{\alpha}^{\dagger}b_{\alpha} \\
\langle a_{\alpha}^{\dagger}a_{\alpha}^{\dagger}a_{\alpha}a_{\alpha}\rangle_{t}
&=&\langle N_{\alpha}^{(1)}(N_{\alpha}^{(1)}-1)\rangle_{t}  \\
\langle b_{\alpha}^{\dagger}b_{\alpha}^{\dagger}b_{\alpha}b_{\alpha}\rangle_{t}
&=&\langle N_{\alpha}^{(2)}(N_{\alpha}^{(2)}-1)\rangle_{t} \\
\langle a_{\alpha}^{\dagger}a_{\alpha}b_{\beta}^{\dagger}b_{\beta}\rangle_{t}
&=&\langle N_{\alpha}^{(1)}\rangle_{t}\langle N_{\beta}^{(2)}\rangle_{t}  \\
\langle N_{\alpha}^{(1)}N_{\beta}^{(1)}\rangle_{t}&=&\langle N_{\alpha}^{(1)}
\rangle_{t}\langle N_{\beta}^{(1)}\rangle_{t}.
\end{eqnarray}
In order to calculate the single particle states $\varphi_{\alpha}$ and 
$\phi_{\alpha}$ that minimizes the free energy $\Omega$ we have to 
calculate the functional derivatives
\begin{eqnarray} 
{\delta \over {\delta \varphi_{\alpha}^{*}({\bf r})}}\int d{\bf r}\tilde 
\Omega (\varphi_{\alpha}^{*}({\bf r}),\phi_{\alpha}^{*}({\bf r}))&=& 0 
\label{func1} \\
{\delta \over {\delta \phi_{\alpha}^{*}({\bf r})}}\int d{\bf r}\tilde 
\Omega (\varphi_{\alpha}^{*}({\bf r}),\phi_{\alpha}^{*}({\bf r})) &=& 0. 
\label{func2}
\end{eqnarray}
For the $\tilde \Omega_{1}$ and $\tilde \Omega_{2}$ parts we can use the 
results for the single condensate \cite{Gold,Huse}. 
From Eqs. (\ref{func1}) and 
(\ref{func2}) we then get the equations
\begin{eqnarray}
&&[-{{\hbar^{2}\over {2m_{1}}}}\nabla^{2}+U_{1}({\bf r})-E_{\alpha}^{(1)}]
\varphi_{\alpha}({\bf r})\langle N_{\alpha}^{(1)}\rangle_{t}+v_{1}\{|\varphi_{
\alpha}({\bf r})|^{2}\langle N_{\alpha}^{(1)}(N_{\alpha}^{(1)}-1) \rangle_{t} 
\nonumber \\ && \\
&&  +\sum_{\beta}\langle N_{\beta}^{(1)}\rangle_{t}|\varphi_{
\beta}({\bf r})|^{2}\langle N_{\alpha}^{(1)} \} \varphi_{\alpha}({\bf r})+
w\sum_{\beta}|\phi_{\beta}({\bf r})|^{2}\langle N_{\beta}^{(2)}\rangle_{t}
\langle N_{\alpha}^{(1)}\rangle_{t}\varphi_{\alpha}({\bf r})=0 \nonumber \\
&& \nonumber \\
&&[-{{\hbar^{2}\over {2m_{2}}}}\nabla^{2}+U_{2}({\bf r})-E_{\alpha}^{(2)}]
\phi_{\alpha}({\bf r})\langle N_{\alpha}^{(2)}\rangle_{t}+v_{2}\{|\phi_{
\alpha}({\bf r})|^{2}\langle N_{\alpha}^{(2)}(N_{\alpha}^{(2)}-1) \rangle_{t} 
\nonumber \\ && \\
&&  +\sum_{\beta}\langle N_{\beta}^{(2)}\rangle_{t}|\phi_{
\beta}({\bf r})|^{2}\langle N_{\alpha}^{(2)}\rangle_{t} \} 
\phi_{\alpha}({\bf r})+
w\sum_{\beta}|\varphi_{\beta}({\bf r})|^{2}\langle N_{\beta}^{(1)}\rangle_{t}
\langle N_{\alpha}^{(2)}\rangle_{t}\phi_{\alpha}({\bf r})=0. \nonumber
\end{eqnarray}
We also keep in mind the difference between the condensed phase and the 
normal phase concerning the single particle averages, see Ref. \cite{us}
\begin{eqnarray}
\langle N_{\alpha}^{(i)}N_{\beta}^{(i)}\rangle_{t}&=&\langle N_{\alpha}^{(i)}
\rangle_{t}
\langle N_{\beta}^{(i)}\rangle_{t} \qquad \alpha \ne  \alpha_{0}, i=1,2 \\
\langle N_{\alpha}^{(i)}(N_{\alpha}^{(i)}-1)\rangle_{t}&=& 2\langle N_{
\alpha}^{(i)} \rangle_{t}^{2} \qquad\qquad \alpha \ne \alpha_{0},i=1,2 \\
\langle N_{\alpha_{0}}^{(i)}(N_{\alpha_{0}}^{(i)}-1)\rangle_{t}&=& N_{0}^{(i)}
(N_{0}^{(i)}-1)\approx {N_{0}^{(i)}}^{2}. 
\end{eqnarray}
We can now have three different regions depending on the parameters: Two 
condensates, a condensed phase and a normal phase or finally two normal 
phases. The three sets of equations are

{\bf a) Two condensates}

The temperature is now below the critical temperature $T_{1}^{c}$ and 
$T_{2}^{c}$ for both gases and we have
\begin{eqnarray}
\left [ -{\hbar^{2} \over {2m_{1}}}\nabla^{2}+U_{1}({\bf r})+
v_{1}(2\rho_{n}^{1}({\bf r})
+\rho_{0}^{1}({\bf r}))+w \Gamma_{1}({\bf r})\right ]
\varphi_{\alpha_{0}}({\bf r})& = &
 E_{\alpha_{0}}^{(1)} \varphi_{\alpha_{0}}({\bf r}) \\
\left [ -{\hbar^{2} \over {2m_{1}}}\nabla^{2}+U_{1}({\bf r})+
2v_{1}(\rho_{n}^{1}({\bf r})+\rho_{0}^{1}({\bf r}))+w \Gamma_{1}({\bf r})
\right ]
\varphi_{\alpha}({\bf r})& = & E_{\alpha}^{(1)} \varphi_{\alpha}({\bf r}) \\
\left [ -{\hbar^{2} \over {2m_{2}}}\nabla^{2}+U_{2}({\bf r})+
v_{2}(2\rho_{n}^{2}({\bf r})
+\rho_{0}^{2}({\bf r}))+w \Gamma_{2}({\bf r})\right ] 
\phi_{\alpha_{0}}({\bf r})& = &
 E_{\alpha_{0}}^{(2)} \phi_{\alpha_{0}}({\bf r}) \\
\left [ -{\hbar^{2} \over {2m_{2}}}\nabla^{2}+U_{2}({\bf r})+
v_{2}(2\rho_{n}^{2}({\bf r})
+\rho_{0}^{2}({\bf r}))+w \Gamma_{2}({\bf r})\right ] \phi_{\alpha}
({\bf r})& = & E_{\alpha}^{(2)} \phi_{\alpha}({\bf r}), 
\end{eqnarray}
where
\begin{eqnarray}
\rho_{n}^{1}({\bf r})&=&\sum_{\alpha\ne\alpha_{0}}\langle N_{\alpha}^{(1)}
\rangle_{t}|\varphi_{\alpha}({\bf r})|^{2} \\
\rho_{0}^{1}({\bf r})&=& N_{\alpha_{0}}^{(1)}|\varphi_{\alpha_{0}}
({\bf r})|^{2} \\
\rho_{n}^{2}({\bf r})&=&\sum_{\alpha\ne\alpha_{0}}\langle N_{\alpha}^{(2)}
\rangle_{t}|\phi_{\alpha}({\bf r})|^{2} \\
\rho_{0}^{2}({\bf r})&=& N_{\alpha_{0}}^{(2)}|\phi_{\alpha_{0}}
({\bf r})|^{2} \\
\Gamma_{1}({\bf r})&=&N_{\alpha_{0}}^{(2)}|\phi_{\alpha_{0}}({\bf r})|^{2}+
\sum_{\alpha\ne\alpha_{0}}\langle N_{\alpha}^{(2)}
\rangle_{t}|\phi_{\alpha}({\bf r})|^{2} \\
\Gamma_{2}({\bf r})&=&N_{\alpha_{0}}^{(1)}|\varphi_{\alpha_{0}}({\bf r})|^{2}+
\sum_{\alpha\ne\alpha_{0}}\langle N_{\alpha}^{(1)}
\rangle_{t}|\varphi_{\alpha}({\bf r})|^{2} \\
\langle N_{\alpha}^{(1)}\rangle_{t}&=&1/(e^{\beta (E_{\alpha}^{(1)}-\mu_{1})}-1) \\
\langle N_{\alpha}^{(2)}\rangle_{t}&=&1/(e^{\beta (E_{\alpha}^{(2)}-
\mu_{2})}-1). 
\end{eqnarray}

{\bf b) A condensed phase and a normal phase}

The temperature is now below one of the critical temperatures. We give here 
the equations with system (1) in the condensed phase,
\begin{eqnarray} 
\left [ -{\hbar^{2} \over {2m_{1}}}\nabla^{2}+U_{1}({\bf r})+
v_{1}(2\rho_{n}^{1}({\bf r})
+\rho_{0}^{1}({\bf r}))+w \Gamma_{1}({\bf r})\right ] 
\varphi_{\alpha_{0}}({\bf r})& = &
 E_{\alpha_{0}}^{(1)} \varphi_{\alpha_{0}}({\bf r}) \label{cn1} \\
\left [ -{\hbar^{2} \over {2m_{1}}}\nabla^{2}+U_{1}({\bf r})+
2v_{1}(\rho_{n}^{1}({\bf r})+\rho_{0}^{1}({\bf r}))+w \Gamma_{1}({\bf r})
\right ]
\varphi_{\alpha}({\bf r})& = & E_{\alpha}^{(1)} \varphi_{\alpha}({\bf r}) 
\label{cn2} \\
\left [ -{\hbar^{2} \over {2m_{2}}}\nabla^{2}+U_{2}({\bf r})+
2v_{2}\rho^{2}({\bf r})+w \Gamma_{2}({\bf r})\right ] 
\phi_{\alpha}({\bf r})& = &
 E_{\alpha}^{(2)} \phi_{\alpha}({\bf r}) \label{cn3}
\end{eqnarray}
with
\begin{eqnarray}
\rho_{n}^{1}({\bf r})&=&\sum_{\alpha\ne\alpha_{0}}\langle N_{\alpha}^{(1)}
\rangle_{t}|\varphi_{\alpha}({\bf r})|^{2} \\
\rho_{0}^{1}({\bf r})&=& N_{\alpha_{0}}^{(1)}|\varphi_{\alpha_{0}}
({\bf r})|^{2} \\
\rho^{2}({\bf r})&=&\sum_{\alpha}\langle N_{\alpha}^{(2)}
\rangle_{t}|\phi_{\alpha}({\bf r})|^{2} \\
\Gamma_{1}({\bf r})&=& \sum_{\alpha}\langle N_{\alpha}^{(2)}
\rangle_{t}|\phi_{\alpha}({\bf r})|^{2} \\
\Gamma_{2}({\bf r})&=&N_{\alpha_{0}}^{(1)}|\varphi_{\alpha_{0}}({\bf r})|^{2}+
\sum_{\alpha\ne\alpha_{0}}\langle N_{\alpha}^{(1)}
\rangle_{t}|\varphi_{\alpha}({\bf r})|^{2}. 
\end{eqnarray}

{\bf c) Two normal phases}

The temperature is here above the critical temperature for both gases.
\begin{eqnarray}
\left [ -{\hbar^{2} \over {2m_{1}}}\nabla^{2}+U_{1}({\bf r})+
2v_{1}\rho^{1}({\bf r})+w \Gamma_{1}({\bf r})\right ] 
\varphi_{\alpha}({\bf r})& = &
 E_{\alpha}^{(1)} \varphi_{\alpha}({\bf r}) \label{nn1} \\
\left [ -{\hbar^{2} \over {2m_{2}}}\nabla^{2}+U_{2}({\bf r})+
2v_{2}\rho^{2}({\bf r})+w \Gamma_{2}({\bf r})\right ] 
\phi_{\alpha}({\bf r})& = &
 E_{\alpha}^{(2)} \phi_{\alpha}({\bf r}) \label{nn2}
\end{eqnarray}
with
\begin{eqnarray}
\rho^{1}({\bf r})&=&\sum_{\alpha}\langle N_{\alpha}^{(1)}
\rangle_{t}|\varphi_{\alpha}({\bf r})|^{2} \\
\rho^{2}({\bf r})&=&\sum_{\alpha}\langle N_{\alpha}^{(2)}
\rangle_{t}|\phi_{\alpha}({\bf r})|^{2} \\
\Gamma_{1}({\bf r})&=& \sum_{\alpha}\langle N_{\alpha}^{(2)}
\rangle_{t}|\phi_{\alpha}({\bf r})|^{2} \\
\Gamma_{2}({\bf r})&=& \sum_{\alpha}\langle N_{\alpha}^{(1)}
\rangle_{t}|\varphi_{\alpha}({\bf r})|^{2}.
\end{eqnarray}

The chemical potentials $\mu_{1}$ and $\mu_2$ are calculated from the 
particle numbers
\begin{eqnarray}
N^{(1)}=\sum_{\alpha}{1\over {e^{\beta (E_{\alpha}^{(1)}-\mu_{1})}-1}} \\
N^{(2)}=\sum_{\alpha}{1\over {e^{\beta (E_{\alpha}^{(2)}-\mu_{2})}-1}}. 
\end{eqnarray}
In the condensed phase we have $\mu_{1}=E_{\alpha_{0}}^{(1)}$ and 
$\mu_{2}=E_{\alpha_{0}}^{(2)}$.

\section{The numerical methods}

The numerical methods have been thouroghly explained in Ref. \cite{us}. We 
are here only going to give the general ideas. 
The equations are highly nonlinear and therefore have to be solved iteratively.
 This means, we start by guessing the densities and solving the ordinary 
differential equations. The resulting eigenvalues and eigenstates are then 
used to obtain new densities. This procedure is then iterated until we have a 
self consistent solution. In the experiments with Bose condensed gases the 
trapping potential has been approximately harmonic. The two gases have 
different external potentials because they are trapped in different spin 
states. This then gives us the two different external potentials
\begin{eqnarray}
U_{1}(x,y,z)&=&{1\over 2} m_{1} (\Omega_{x_{1}}^{2}x^{2}+
\Omega_{y_{1}}^{2}y^{2}+\Omega_{z_{1}}^{2}z^{2}) \\
U_{2}(x,y,z)&=&{1\over 2} m_{2} (\Omega_{x_{2}}^{2}x^{2}+
\Omega_{y_{2}}^{2}y^{2}+\Omega_{z_{2}}^{2}z^{2}). 
\end{eqnarray}
Because of the harmonic external potentials, a natural approach would be to 
expand the solutions in the harmonic oscillator eigenfunctions
\begin{equation}
\varphi ({\bf r})=\sum_{i}a_{i}\Phi_{i}^{HO}. 
\end{equation}
The problem then reduces to an eigenvalue problem for the expansion 
coefficients $a_i$. In the actual calculations we have at most four 
coupled equations that have to be solved simultaneously. This is not a 
problem since we have basically the same situation as in the single 
condensate case. The only drawback is the computing time which is doubled 
compared to the calculations in Ref. \cite{us}.

The method of expansion in harmonic oscillator eigenfunctions is especially 
good in asymmetric environments. However, if we have a spherically symmetric 
geometry, we can easily discretize our solutions on a lattice, 
$\varphi (r)\rightarrow 
\varphi_{i}$, and the derivatives turn into differences, 
which then gives us a tridiagonal 
eigenvalue problem whose solutions directly gives the desired quantities. This 
method works best in situations where the problem is effectively one 
dimensional.

\section{Results}

In this paper we calculate the density of the two-component condensate in a 
harmonic trap in two and three dimensions including asymmetry in the two 
dimensional case. We also present a phase diagram for a three dimensional 
spherically symmetric trap. The equations are put into dimensionless form 
by the scaling 
\begin{equation}
r={\hbar^{1\over 2} \over {[ m_{1}m_{2}\Omega_{1}\Omega_{2}]^{1\over 4}}} r'  
\end{equation}
and $x=\lambda x', y=\lambda y'$ with 
\begin{equation}
\lambda={\hbar^{1\over 2} \over {[m_{1}m_{2}]^{1\over 2}[\Omega_{x_{1}}
\Omega_{x_{2}}\Omega_{y_{1}}\Omega_{y_{2}}]^{1\over 8}}}.
\end{equation}
This gives us the dimensionless energies scaled to $
{1\over 2}\hbar (\Omega_{x_{1}}\Omega_{x_{1}}\Omega_{y_{1}}
\Omega_{y_{2}})^{1/4} \sqrt{m_{2}/m_{1}}$ and $
{1\over 2}\hbar (\Omega_{x_{1}}\Omega_{x_{1}}\Omega_{y_{1}}
\Omega_{y_{2}})^{1/4} \sqrt{m_{1}/m_{2}}$.

\subsection{The density calculations}

The most natural thing to calculate from the Hartree-Fock equations are the 
densities. In the spherically symmetric case we have the external potentials
\begin{eqnarray}
U_{1}(r)&=&{1\over 2} m_{1} \Omega_{1}^{2} r^{2} \\
U_{2}(r)&=&{1\over 2} m_{2} \Omega_{2}^{2} r^{2}. 
\end{eqnarray}
Throughout the calculations in this paper we have chosen the ratio between 
$m_1$ and $m_2$ to be $m_{1}/m_{2}=1$ and the interactions $v_1$ and $v_2$ 
to be $v_{1}=0.02$ and $v_{2}=0.01$. In Fig.1 we show the spherically symmetric
 density for the two condensates with $N=1000$ particles in both gases and the 
interaction strength between the different particles is put to $w=0.012$. Here 
we can see that the two different particles hardly know anything about each 
other. The inverse temperature is here $\beta =0.3$. In Fig.2 we
have incresed the interaction strength $w$ to $w=0.05$. The condensate of the 
more weakly interacting particles is pushed away from the center of the trap 
and is forming a shell around the particles in the center of the trap. 
These results were obtained by spherically symmetric eigensolutions which 
means that we can not say anything about the existance of states with lower 
energy that could posess an asymmetric geometry. This suggests to
 look for a solution of an asymmetric two-component condensate.

It is extremely time and memory consuming to solve the full three dimensional 
asymmetric case. We therefore concentrate on the two dimensional two-component 
condensate. We then have the external potentials
\begin{eqnarray}
U_{1}(x,y)&=&{1\over 2}m_{1}(\Omega_{x_{1}}^{2}x^{2}+\Omega_{y_{1}}^{2}y^{2})\\
U_{2}(x,y)&=&{1\over 2}m_{2}(\Omega_{x_{2}}^{2}x^{2}+\Omega_{y_{2}}^{2}y^{2}).
\end{eqnarray}
In the two dimensional calculations we have chosen the asymmetry of the trap 
to be 
\begin{equation}
{\Omega_{x_{i}}\over \Omega_{y_{i}}}=\sqrt{8} \qquad i=1,2
\end{equation}
with $\Omega_{\alpha_{1}}/\Omega_{\alpha_{2}}=\sqrt{2}$.
In Figs. 3)-5) we show the density of the two condensates at the inverse 
temperature $\beta=1.0$. If we start the iteration by putting the initial 
densities at for instance $x=1,y=1$ and $x=1,y=-1$, we end the iteration with 
two separated 
condensates aligned in the weaker trap direction. In Fig.3 we have 
used $w=0.1$ 
and in Fig.4 we have incresed the interaction to $w=0.3$, 
where we can see that 
the two condensates get further pushed away from each other with incresing 
interaction strength. However, if we start by putting both condensates at the 
center of the trap, we get a stable solution that is symmetric also in the 
weaker 
trap direction and does not show a separation into two displaced condensates.
 This is shown in Fig.5. 
The ground state energy for this case is higher than in the purely asymmetric 
situation which suggests that the spontaneous symmetry breaking occures.
This is 
more clearly seen if we look at the two dimensional rotationally 
symmetric potential. 
Starting the iteration with the initial densities away from the center, gives 
us two condensates separated into two distinct peaks. 
This is shown in Fig.6 with 
$\beta=1.0$ and $w=0.1$. In Fig.7 we have the same situation but now we start 
the iteration with the densities at the center of the trap. 
The calculation
converges nicely and the stable solution shows two condensate peaks sitting on 
top of each other. In this case the interaction strength is not strong enough 
to create a ring of the lighter atoms around the center, which was seen in the 
three dimensional calculations in Fig.2. 

\subsection{The phase diagram}

The Hartree-Fock equations do not describe exactly the 
transition region between
an ordinary gas and a Bose condensed one, but they do suggest 
a general view of
what is going on. If we define the critical temperature as the temperature
where the number of particles in the condensate goes to zero, we can calculate
this number of particles with the equations for $(T<T_c)$ at
different temperatures until we reach the point where all particles in
the condensate have been depleted. In Fig.8 we show three phase diagrams as 
function of temperature and interaction strength $w$, with a fixed number of 
particles $N^{(1)}=100$ and $N^{(2)}$ varying between 100 and 728. We see that 
three regions need to be covered. We first solve the equations for two 
condensates with a fixed $w$ and increse the temperature until one of 
the condensates disappears, where we find our phase transition. Incresing the 
temperature even more means that we are in the one-condensed-one-normal-phase 
region and we have to use Eqs. (\ref{cn1})-(\ref{cn3}). Incresing the 
temperature further, finally destroys the remaining condensate and we have 
found our 
second transition point. This kind of calculation has been repeted for 
different interaction strengths and particle numbers. Adjusting the particle 
numbers so that the two transition lines almost coincide, 
we find that there exists 
a region where one can go from two condensates to normal gases by lowering 
the temperature. This situation is shown in Fig.8b. The phenomena exhibited 
in our phase diagrams may not give an accurat picture of the real systems, but 
they can be believed to suggest the trends expected in the actual experiments.

\section{Conclusions}

In this paper we have, for the first time, presented density calculations of a 
two-component condensate as functions of temperature. We have found a 
symmetry 
breaking in the two dimensional case,
 where the two-component condensate sitting at 
the center of the trap on top of each other posess a higher energy than the 
situation where the two condensates have separated  and form two individual 
peaks. This is a case 
of spontaneous symmetry breaking. We also present a phase diagram 
which describes the different regions 
with a two-component condensate, one single condensate and finally no 
condensate as 
function of temperature and the interaction strength between the two different 
atoms. The phase diagram has been presented for three different pairs of 
particle numbers and calculated in three dimensions with 
spherical symmetric external potentials. This means that we can not see the 
symmetry breaking which we saw in the two dimensional calculations which, in 
fact, gave an asymmetric solution in the symmetrical situation. 
The Hartree-Fock equations do not describe the transition region 
well because they neglect all critical fluctuations. They do, however, give us 
an idea what may be happening around the transition point. At some critical 
particle number we find a region where it is possible by changing the 
temperature at fixed interaction strengths $w$, 
to go from a two-component condensate to two 
noncondensed gases. 

In these calculations, we have only used a few hundred 
particles. To increse the number of particles to realistic values 
($\sim 10^6$) is beyond the capacity of available computers. So is a full 
treatment of totally asymmetric three dimensional traps. With this small 
number of particles, the interaction strength, $w$, needs to be chosen 
unrealistically large to bring out the observed features. 
The small number of particles may give rise to finite size 
effects as for 
instance regions with an inconsistant fraction of condensate particles 
calculated from the two-component condensate equations and the single 
condensate 
equations. With incresing particle number, we may expect the inconsistensy to 
disappear and give a 
crossing from a two-component condensate into a phase with two normal 
components for all interaction strengths $w$. 

The numerics was performed with a grid method in the three dimensional 
spherically symmetric case. This is a very stable and fast method. The only 
drawback is that it can in 
practice only be used in situations where the problem is
 effectively reduced to a one dimensional. 
The expansion method, on the other hand, 
is very fragile in the two dimensional calculations, and great care has to be 
taken in order not to get runaway iterations which do not converge to 
physically stable solutions.

Finally we want to speculate on some features observed in our numerical 
computations. They may be due only to shortcomings of the numerical approach 
or limitations of the HF method, but they point to interesting 
possibilities in the behaviour of the real systems.

First we look at situations like in Fig.8b. There the two-condensate 
and no-condensate boundaries are very close to each other; they may, in fact, 
be found to cross. In such regions, we find that the critical temperatures 
found from the two-condensate and from the single-condensate sides are 
inconsistent. If we could trust the HF-calculations, this may indicate a 
hysteresis signalling a change to a first order transition for one of the 
components. No such conclusion can be proven from the HF-approach, but it 
points to the possibility that the change of one component can qualitatively 
affect the behaviour of the other one even to the extent that its order may 
change.

Secondly, the use of too few states in the calculations gives a distorted and 
asymmetric solution. In two dimensions and for not too high temperatures 
$(\beta=1.0)$, we get smooth densities for about ten states in each direction. 
Because of the unrealistically large values for $w$ used, the two condensates 
repell each other strongly. This shows up as a drift instability which 
eventually develops into an oscillational instability in the solution, if the 
computer iteration is continued well after a stable solution is found. It 
seems that this can be overcome by incresing the number of states involved. 
The physical contents of our numerical observation may be that it signals the 
break up of a solution only locally stable. Thus we find it to occur much 
more readily for the symmetric situation, whereas the asymmetric solutions 
are much more stable.

We have thus found our numerics to indicate the change of order of the 
transition and the instability of some locally stable solutions owing to the 
interaction between the two condensates. Admittedly neither our numerical 
method nor our theoretical formulation (HF) allows any claims to the reality 
of the effects. They do, however, offer challenging possibilities for 
further experimental and theoretical investigations.

Unfortunately, it seems to be difficult to approach these problems from the 
two-component Gross-Pitaevskii equations. As these omit the
atoms above the condensate, it is difficult to compute the effects on one 
condensate by the properties of the other one. On the other hand, any phase 
transition theory superior to the HF-approach appears to offer 
unsurmountable computational difficulties. What progress can be achieved 
on these difficult questions remains to be seen.

\newpage

\begin{center}
FIGURE CAPTIONS
\end{center}

Figure 1: The spherically symmetric density of the two condensates. The two 
condensates hardly see each other at $w=0.012$. The insert shows the excited
 part. The temperature was here $T=3.33$ and the interactions $v_{1}=0.02$
 and $v_{2}=0.01$ with $N^{(1)}=1000$ and $N^{(2)}=1000$. 
\vskip 1cm
Figure 2: The same situation as in Fig.1 with a stronger mutual 
interaction $w=0.05$.
Here we can see that the $N^{(2)}$-particles are pushed away from the center.
\vskip 1cm
Figure 3: The two dimensional asymmetric case were the condensates have
formed two separated peaks around the center of the trap. The interaction is
here $w=0.1$ with $\Omega_{x}/\Omega_{y}=\sqrt{8}$, $\Omega_{\alpha_1}/
\Omega_{\alpha_2}=\sqrt{2}$ and the temperature 
$T=1.0$. The iteration is here started away from the center with two Gaussian 
condensate densities at (1,1) and (1,-1). The ground state energies are here
 $E_{0}^{(1)}=3.813$ and $E_{0}^{(2)}=2.924$.
\vskip 1cm
Figure 4: The same situation as in Fig. 3 with the stronger 
interaction $w=0.3$. The 
two condensates are pushed further away from each other because of their 
stronger repulsive interaction.   
\vskip 1cm
Figure 5: The same situation as in Fig.3. The iteration is here 
started with the condensates in the 
center of the trap with $w=0.1$ and $T=1.0$. The condensates sit on top of 
each other. The ground state energy is 
higher than in the asymmetric formation, with $E_{0}^{(1)}=4.114$ and 
$E_{0}^{(2)}=3.150$; cf the values given in Fig. 3.
\vskip 1cm
Figure 6: The two dimensional spherically symmetric situation were we find 
two separated density peaks with the ground state energies 
$E_{0}^{(1)}=3.80$, $E_{0}^{(2)}=2.97$ 
and $w=0.1$. The different lines show the 
density contours as indicated on the left side of the figure. The 
iteration is started away from the center.
\vskip 1cm
Figure 7: The two condensates situated in the center of the trap on top of each
other. The iteration is started with two Gaussian densities positioned at  
the center of the trap. The ground state energy in this case is found to be 
slightly higher than 
the off-center iteration in Fig.6. The energies are here $E_{0}^{(1)}=3.81$ 
and $E_{0}^{(2)}=2.98$.
\vskip 1cm
Figure 8: The phase diagram described for three different pairs of particle 
number. In a) we have $N^{(1)}=100$ and $N^{(2)}=100$. Here we can see that 
the 
transition line between region II and III (one condensate and no condensate) 
is not sensitive to changes in $w$. In region II the 
$N^{(1)}$-particles are condensed and the $N^{(2)}$-particles are in the 
normal phase. In b) we have $N^{(1)}=100$ and 
$N^{(2)}=617$, 
and we find a region where we have a transition between a 
two-component condensate (I) and 
no condensate (III). In c) we have incresed the relative particle number to 
$N^{(1)}=100$ and $N^{(2)}=728$. Region I with the two-component condensate is 
strongly 
supressed as a function of $w$, whereas the transition between one condensate 
and no condensate (II and III) is not sensitive to changes in $w$. In this 
case, for region II, the $N^{(2)}$-particles are condensed and the 
$N^{(1)}$-particles are in the normal phase

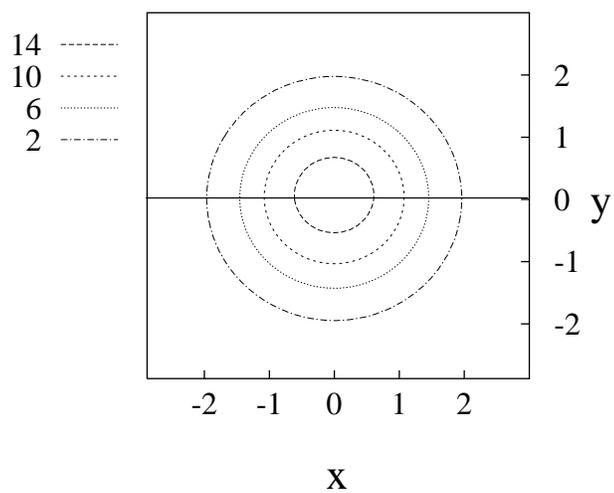
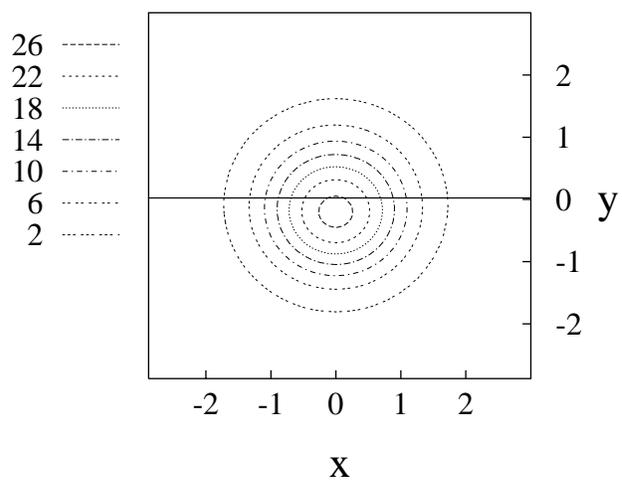
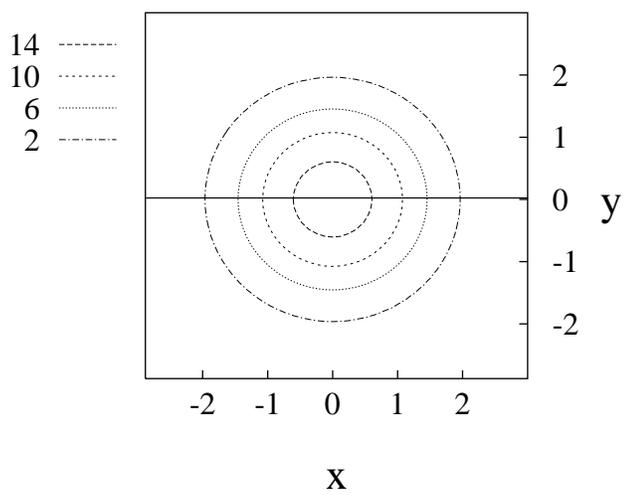
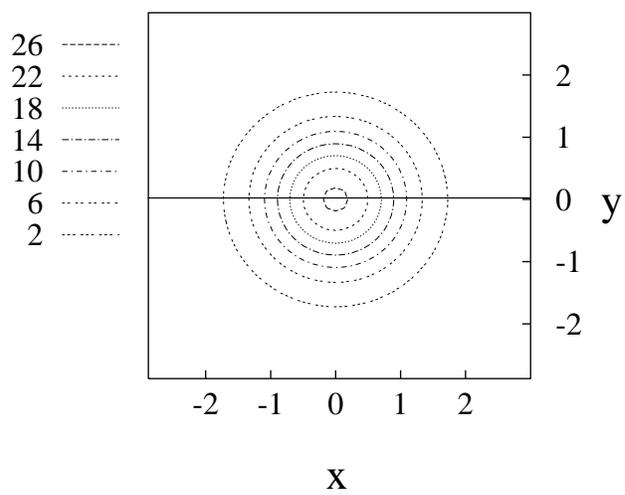
\end{document}